\documentclass[12pt,a4paper]{article}

\usepackage{amsmath,amssymb}
\usepackage{verbatim}
\usepackage{graphicx}
\usepackage[nosort]{cite}
\usepackage[hyperref,bulletsep]{collect}

\makeatletter \@addtoreset{equation}{section} \makeatother

\makeatletter
\let\old@startsection=\@startsection
\renewcommand{\@startsection}[6]{\old@startsection{#1}{#2}{#3}{#4}{#5}{#6\mathversion{bold}}}
\makeatother

\makeatletter
\let\old@makecaption=\@makecaption
\def\@makecaption{\small\old@makecaption}
\makeatother

\let\oldPhi=\Phi
\let\oldPsi=\Psi
\let\oldGamma=\Gamma
\let\oldDelta=\Delta
\let\oldSigma=\Sigma
\let\oldTheta=\Theta
\let\oldPi=\Pi
\renewcommand{\Phi}{\mathnormal{\oldPhi}}
\renewcommand{\Psi}{\mathnormal{\oldPsi}}
\renewcommand{\Gamma}{\mathnormal{\oldGamma}}
\renewcommand{\Sigma}{\mathnormal{\oldSigma}}
\renewcommand{\Delta}{\mathnormal{\oldDelta}}
\renewcommand{\Theta}{\mathnormal{\oldTheta}}
\renewcommand{\Pi}{\mathnormal{\oldPi}}

\newcommand{\superN}{\mathcal{N}}

\newcommand{\ham}{\mathcal{H}}
\newcommand{\gym}{g\indups{YM}}
\newcommand{\Op}{\mathcal{O}}
\newcommand{\Tr}{\mathop{\mathrm{Tr}}}
\newcommand{\diag}{\mathop{\mathrm{diag}}}

\newcommand{\fld}[1]{\mathcal{#1}}

\newcommand{\Integers}{\mathbb{Z}}

\ifx\genfrac\sdflkaj

\else

\fi
\newcommand{\sfrac}[2]{{\textstyle\frac{#1}{#2}}}

\newcommand{\ihalf}{\sfrac{i}{2}}

\newcommand{\indup}[1]{_{\mathrm{#1}}}
\newcommand{\indups}[1]{_{\mathrm{\scriptscriptstyle #1}}}

\newcommand{\rep}[1]{{\mathbf{#1}}}

\newcommand{\alg}[1]{\mathfrak{#1}}
\newcommand{\grp}[1]{\mathrm{#1}}

\newcommand{\lrbrk}[1]{\left(#1\right)}
\newcommand{\bigbrk}[1]{\bigl(#1\bigr)}

\newcommand{\acomm}[2]{\{#1,#2\}}

\newcommand{\set}[1]{\{#1\}}
\newcommand{\state}[1]{\mathopen{|}#1\mathclose{\rangle}}

\newcommand{\orbi}{\oldGamma}

\newcommand{\nn}{\nonumber}
\newcommand{\nln}{\nonumber\\}

\newcommand{\earel}[1]{\mathrel{}&\hspace{-2\arraycolsep}#1\hspace{-2\arraycolsep}&\mathrel{}}
\newcommand{\eq}{\earel{=}}

\def\[{\begin{equation}}
\def\]{\end{equation}}
\def\<{\begin{eqnarray}}
\def\>{\end{eqnarray}}

\makeatletter
\def\mr@ignsp#1 {\ifx\:#1\@empty\else #1\expandafter\mr@ignsp\fi}%
\newcommand{\multiref}[1]{\begingroup
\xdef\mr@no@sparg{\expandafter\mr@ignsp#1 \: }%
\def\mr@comma{}%
\@for\mr@refs:=\mr@no@sparg\do{\mr@comma\def\mr@comma{,}\ref{\mr@refs}}%
\endgroup}
\makeatother

\newcommand{\hypref}[2]{\ifx\href\asklfhas #2\else\href{#1}{#2}\fi}

\newcommand{\secref}[1]{Sec.~\multiref{#1}}

\newcommand{\figref}[1]{Fig.~\multiref{#1}}
\renewcommand{\eqref}[1]{(\multiref{#1})}

\ifx\href\asklfhas\newcommand{\href}[2]{#2}\fi

\newcommand{\vect}[1]{\mathbf{#1}}

\begin{document}
\setcounter{page}{0}

\thispagestyle{empty}
\begin{flushright}\footnotesize
\texttt{hep-th/0510209}\\
\texttt{PUTP-2179}\\
\texttt{KITP-NSF-05-88}
\end{flushright}
\vspace{0.5cm}

\renewcommand{\thefootnote}{\fnsymbol{footnote}}
\setcounter{footnote}{0}

\begin{center}
{\Large\textbf{\mathversion{bold}%
The Bethe Ansatz for $\Integers_S$ Orbifolds\\
of $\superN=4$ Super Yang-Mills Theory%
}\par}
\vspace{1cm}

\textsc{N.~Beisert$^a$ and R.~Roiban$^{b}$}
\vspace{5mm}

\textit{$^{a}$ Joseph Henry Laboratories\\
Princeton University\\
Princeton, NJ 08544, USA} \vspace{3mm}

\textit{$^{b}$ Physics Department\\
Pennsylvania State University\\
University Park, PA 16802, USA}
\vspace{3mm}

\texttt{nbeisert@princeton.edu}\\
\texttt{radu@phys.psu.edu}
\par\vspace{1cm}

\vfill

\textbf{Abstract}\vspace{5mm}

\begin{minipage}{12.7cm}
Worldsheet techniques can be used to argue for the integrability of
string theory on $AdS_5\times S^5/\mathbb{Z}_S$, which is dual to
the strongly coupled $\mathbb{Z}_S$-orbifold of $\mathcal{N}=4$ SYM.
We analyze the integrability of these field theories in the
perturbative regime and construct the relevant Bethe equations.
\end{minipage}

\vspace*{\fill}

\end{center}

\newpage
\setcounter{page}{1}
\renewcommand{\thefootnote}{\arabic{footnote}}
\setcounter{footnote}{0}

\section{Introduction}

Our understanding of the AdS/CFT correspondence benefited greatly
from the discovery of integrable structures both on the gauge theory
\cite{Minahan:2002ve,Beisert:2003tq,Beisert:2003yb,Beisert:2003ys}
and string theory side \cite{Mandal:2002fs,Bena:2003wd,Kazakov:2004qf}
(see \cite{Beisert:2004ry,Beisert:2004yq,Zarembo:2004hp,Plefka:2005bk} for reviews). 
A natural question to ask is to what extent we can deform the
model while preserving full integrability. 
This should lead to a better understanding of the integrable structures
in large-$N$ field theories in general and $\superN=4$ Super Yang-Mills (SYM) 
in particular.

Soon after the AdS$_{d+1}$/CFT$_d$ correspondence was formulated
\cite{Maldacena:1998re,Gubser:1998bc,Witten:1998qj},
it was realized that modding out by a discrete
subgroup of the R-symmetry group $\grp{SU}(4)$ 
leads to candidate dual pairs with reduced
supersymmetry \cite{Kachru:1998ys,Lawrence:1998ja}.
The properties of the orbifold group
determine the amount of preserved (super)symmetry. From the perspective
of a four-dimensional theory,  if the discrete orbifold group
$\orbi\in\grp{SU}(2)$, then one finds a $\superN=2$ superconformal
gauge theory, while $\orbi\in \grp{SU}(3)$ produces a $\superN=1$
superconformal gauge theory. For all other orbifold groups $\orbi$ the
supersymmetry is completely broken. Depending on the precise
embedding $\orbi\subset \grp{SU}(4)$ some bosonic symmetries may
survive the orbifold projection.

Similar to the states of string theory orbifolds,
the operators in orbifold
field theories are organized in representations of the orbifold
group. Evidence was provided in \cite{Bershadsky:1998mb,Bershadsky:1998cb} that all
correlation functions of untwisted operators (i.e.~the operators
that do not transform under the quantum symmetry $\orbi$)
coincide in the planar limit with the correlation functions
in the parent $\superN=4$ SYM theory. Consequently, up to a trivial
rescaling $\lambda\mapsto \lambda/|\orbi|$, the anomalous dimensions
of untwisted operators in the orbifold theory are the same as the
parent $\superN=4$ SYM theory. The anomalous dimensions of twisted
operators do not obey such an inheritance principle and must be computed
separately. As long as the orbi\-fold quantum symmetry is
unbroken%
\footnote{Non-supersymmetric orbifold theories exhibit certain
instabilities \cite{Tseytlin:1999ii,Dymarsky:2005uh}
which may be interpreted in terms of the condensation of
closed string tachyons \cite{Adams:2001jb,Dymarsky:2005nc}. 
In the condensation process the quantum
symmetry of the orbifold is spontaneously broken potentially
leading to a more
complicated structure of the dilatation operator.}
operators belonging to different twisted sectors do not mix under
renormalization group flow 
and the anomalous dimensions of operators in a fixed twisted
sector can be computed independently.

In this note we set up the Bethe ansatz for general
abelian orbifolds, as a deformation of the Bethe ansatz for the parent
theory. We begin by discussing in detail general $\Integers_S$
orbifolds, explicitly computing the anomalous dimensions for
several classes of operators. Examples of anomalous dimensions of
such orbifolds of $\superN=4$ SYM have already appeared in the
literature in the context of plane wave orbifolds \cite{Alishahiha:2002ev,Kim:2002fp}
as well as in connection with the integrability of the dilatation
operator in particular sectors of the theory
\cite{Wang:2003cu,Ideguchi:2004wm}. 
We will compare them with the proposed Bethe ansatz. 
We then extend these results to the general
case. We close with a discussion on further deformations of orbifolds of
$\superN=4$ SYM as well as orbifolds of other deformations of $\superN=4$ SYM.

\section{Orbifolds of Field Theory}

We start by reviewing the construction of orbifolds
for gauge field theories and set up our notation.
For these models we find the action
of the planar dilatation generator and apply it
to find a few sets of anomalous dimensions.

\subsection{Fields}

We consider field theories dual to superstrings
on $AdS_5\times (S^5/\orbi)$ where $\orbi$ is a discrete
subgroup of $\grp{SU}(4)$.
In the gauge theory dual, the orbifold group $\orbi$
acts on the gauge group as well as on the
$\grp{SU}(4)$ R-symmetry group. Consistency of stringy
construction of such field theories requires
that we pick regular representations of the orbifold group 
\cite{Bershadsky:1998mb}.
The field $X$ transforms under the action of
an element of $\orbi$ as follows
\[\label{eq:OrbifoldGeneric}
X\mapsto\gamma (R_\gamma X) \gamma^{-1}=X.
\]
Here $\gamma$ is the representation of the element
as a matrix of the gauge group and
$R_\gamma$ is the representation
in $\grp{SU}(4)$. The orbifold group acts trivially and therefore
\eqref{eq:OrbifoldGeneric} is a constraint for the fields $X$.

In particular, we are interested in a $\orbi=\Integers_S$ orbifold of
$\superN=4$ SYM with gauge group $\grp{U}(SN)$.
The orbifold theory has residual $\grp{U}(N)^S$ local symmetry.
We introduce a $SN\times SN$ matrix $\gamma$
representing a generator of $\Integers_S$.
Using $N\times N$ blocks it has the form
\<\label{eq:OrbifoldMatrix}
\gamma\eq\diag \bigbrk{1,e^{2\pi i/S},e^{4\pi i/S},\ldots,e^{-2\pi i/S}}
\nln\eq
\diag \bigbrk{1,\omega,\omega^2,\ldots,\omega^{S-1}},
\qquad\qquad\qquad
\omega=e^{2\pi i/S}.
\>
A field $X$ of the orbifold theory with definite charges of $\alg{su}(4)$
is defined by the constraint
\[\label{eq:OrbifoldAction}
X
=\exp(2\pi i s_X/S)\, \gamma X \gamma^{-1}
=\omega^{s_X}\, \gamma X \gamma^{-1}.
\]
Alternatively, we can project the $\superN=4$ SYM field $X^{\superN=4}$
to the orbifold field $X$ by means of
\[\label{eq:OrbifoldProject}
X=\frac{1}{S}\sum_{k=1}^{S} \exp(2\pi i ks_X/S)\,\gamma^{k} X^{\superN=4} \gamma^{-k}.
\]
The $\Integers_S$ integer $s_X$ selects which of the (secondary)
diagonals of the $SN\times SN$ matrix $X$ is occupied.
The phase $s_X$ is coupled to the internal $\alg{su}(4)$ symmetry
as follows
\[\label{eq:OrbifoldPhase}
s_X=\vect{t}\cdot \vect{q}_X.
\]
The vector $\vect{q}_X$ represents the $\alg{su}(4)$ charges
of the field $X$ and the vector $\vect{t}$
contains the parameters of the orbifold.
As $\alg{su}(4)$ has rank three, there are three
independent parameters $t_1,t_2,t_3$. These
are taken to be integers and only their
remainder modulo $S$ is relevant.
Let us define $s_X$ through the action on a spinor of $\alg{so}(6)$.
The phases for the four components of the spinor shall be given by
\[\label{eq:OrbifoldSpinor}
\vect{s}\indup{\rep{4}}=(-t_1, t_1-t_2, t_2-t_3, t_3).
\]
Vectors of $\alg{so}(6)$ are given by bi-spinors and conjugate
spinors by triple spinors; their phases $\vect{s}_{\rep{6}}$ and
$\vect{s}_{\rep{\bar 4}}$ can therefore be obtained by adding the corresponding
phases from $\vect{s}_{\rep{4}}$.
Fields invariant under $\alg{su}(4)$ have no phases and
consequently they are block-diagonal, such as the gauge field $\fld{A}_\mu$.
For $\superN=1$ supersymmetric theories,
it is useful to split up $\rep{6}=\rep{3}+\rep{3}$
and employ a $\alg{su}(3)$ notation for the fields.
Then the triplet of complex scalars is related to the
scalar as a bi-spinor as $\phi_k=\Phi_{k4}$
and the corresponding weights are
\[\label{eq:OrbifoldSU3}
\vect{s}\indup{\rep{3}}=(-t_1+t_3, t_1-t_2+t_3, t_2).
\]
%

\subsection{Spin Chains}

A generic local operator invariant under $\grp{U}(N)^S$ is given by
traces of products of the fields $\fld{W}$ and the $\Integers_S$
generator $\gamma$.
Due to the orbifold identity \eqref{eq:OrbifoldAction},
which can be reformulated as
\[\label{eq:OrbifoldShift}
 X \gamma=\exp(2\pi i s_X/S)\, \gamma X ,
\]
it is possible to collect all $\Integers_S$ generators at one point
within a trace.
Single-trace operators are therefore spanned by
\[\label{eq:SpinChainOperator}
\Op=\Tr \gamma^T\fld{W}_{A_1}\fld{W}_{A_2}\ldots\fld{W}_{A_L}.
\]
Here $\fld{W}\in \set{\fld{D}^n\Phi,\fld{D}^n\Psi,\fld{D}^n\fld{F}}$
is a multiple derivative of one of the fields of the theory.
Some of the operators \eqref{eq:SpinChainOperator} vanish
identically. Using the equation \eqref{eq:OrbifoldShift} to commute
one $\gamma$ past all the fields in the operator leads to the conclusion
that the necessary and sufficient condition for non-vanishing operators
is that their total $\Integers_S$ charge is zero,
i.e.:
\[
\label{eq:condition}
\frac{1}{S}\sum_{k=1}^L s_{A_k}\in\Integers~.
\]
Up to the cyclicity of the trace, this local operator is isomorphic
to some spin chain state
\[\label{eq:SpinChainState}
\state{\Op}=\state{\gamma^T;A_1,A_2,\ldots,A_L}.
\]
In this picture, the gauge theory generator of anomalous dimensions
maps to the spin chain Hamiltonian $\ham$.
The energy eigenvalues $E$ of $\ham$ are related to the
anomalous dimensions of local operators by
\[\label{eq:SpinChainEnergy}
\delta D_{\Op}=\frac{\gym^2 N}{8\pi2}\,E_{\state{\Op}}.
\]
By investigating Feynman diagrams, it can be seen that
the $L$-loop Hamiltonian commutes with the orbifolding
procedure in the large-$N$ limit for all operators of length strictly
larger than $L$ \cite{Bershadsky:1998cb}.%
\footnote{A further subtlety arises for non-supersymmetric orbifold
actions, as quantum corrections require the modification of
the tree-level action by certain bilinears in twisted dimension two
operators. The planar anomalous dimensions of length $L$ operators
cannot be not affected by such deformations before $(L-1)$ loops.
In the following we will largely ignore this subtlety.}$^,$%
\footnote{It is not immediately clear whether the orbifold procedure
commutes with the $L$-loop spin chain Hamiltonian for operators of length
smaller than $L$. It is intriguing that this apparent problem sets in
at the same step at which the wrapping interactions start being
relevant \cite{Beisert:2004hm} 
(see also \cite{Ambjorn:2005wa} for recent investigations of this effect).
It would be interesting to establish whether a connection
exists between these two effects and if their mutual consistency
constrains the form of the latter. }
This means that the Hamiltonian for $\superN=4$ SYM is the same
as for the orbifolded theory when there are no
$\Integers_S$ generators involved. For example,
if the nearest-neighbor Hamiltonian for $\superN=4$ SYM
at the leading, one-loop level is given by
\[\label{eq:SpinChainHamN4}
\ham^{\superN=4}_{[12]}\,\fld{W}^{\superN=4}_{A}\fld{W}^{\superN=4}_{B}=
\ham_{AB}^{CD}\,\fld{W}^{\superN=4}_{C}\fld{W}^{\superN=4}_{D}.
\]
Then the action of the orbifolded Hamiltonian yields
\[\label{eq:SpinChainHam}
\ham_{[12]}\,\fld{W}_{A}\fld{W}_{B}=
\ham_{AB}^{CD}\,\fld{W}_{C}\fld{W}_{D}.
\]
%
If $\Integers_S$ generators are present in the interaction region
they should first be shifted away using equation
(\ref{eq:OrbifoldShift}), e.g.
\[\label{eq:SpinChainHamTwist}
\ham_{[12]}\,\fld{W}_{A}\gamma^k\fld{W}_{B}=
\exp(2\pi i k s_A/S) \ham_{[12]}\,\gamma^k\fld{W}_{A}\fld{W}_{B}
=
\exp(2\pi i k s_A/S) \ham_{AB}^{CD}\,\gamma^k\fld{W}_{C}\fld{W}_{D}.
\]
Clearly, the number of $\Integers_S$ generators, i.e.~the
$\Integers_S$ charge $T$ of the local operator is preserved by
$\ham$. Therefore the states \eqref{eq:SpinChainState}
with fixed $T$ form a so-called $T$-twisted sector of the model.

\subsection{Anomalous Dimensions}

It is straightforward to use the description above to compute
the anomalous dimensions of one-excitation states
\[
\Op=\Tr (\gamma^T \fld{W} \fld{Z}^{L-1})
\]
where $\fld{Z}$ is one of the complex scalars
and $\fld{W}$ can be any of the fundamental fields of the theory except
for the gauge field.
The orbifold phases (weights) of the fields
$\fld{Z},\fld{W}$ are assumed to be $s_{\fld{Z}},s_{\fld{W}}$,
respectively. In general, for an operator to be non-trivial it is
necessary to accumulate a trivial phase as $\gamma$ is moved past all the
fields using \eqref{eq:OrbifoldShift}. In the case at hand
\eqref{eq:condition}
reduces to
\[\label{eq:singleexconstraint}
\frac{(L-1)s_{\fld{Z}}+s_{\fld{W}}}{S}\in\Integers.
\]
For those states, the
energy $E$, alias the one-loop planar anomalous dimension 
$\delta D=(\gym^2N/8\pi^2)E$,
is
\[\label{eq:singleex}
E=4\sin^2\frac{\pi s_{\fld{Z}}T}{S}\,.
\]

Given the simplicity of the states of length two, it is also find
relatively closed forms for the anomalous dimensions of all operators
of this length descending from the $\alg{so}(6)$ sector.
Only the operators $\Tr(\gamma^T\acomm{\bar \phi_i}{\phi^i})$ exist
for a general abelian orbifold group.  Their anomalous
dimensions are given by the eigenvalues of
the mixing matrix
\begin{eqnarray}
E=
\left(
\begin{array}{rrr}
2+2\sin^2(\pi s_1T/S) & 2-2\sin^2(\pi s_1T/S) & 2-2\sin^2(\pi s_1T/S)
\\
2-2\sin^2(\pi s_2T/S) & 2+2\sin^2(\pi s_2T/S) & 2-2\sin^2(\pi s_2T/S)
\\
2-2\sin^2(\pi s_3T/S) & 2-2\sin^2(\pi s_3T/S) & 2+2\sin^2(\pi s_3T/S)
\end{array}
\right)~,
\end{eqnarray}
which are relatively complicated functions of the orbifold weights
$\vect{s}_{\rep{3}}=(s_1,s_2,s_3)$ in the $\rep{3}$ representation.
If two weights are equal, $\vect{s}_{\rep{3}}=(2s',s,s)$,
the eigenvalues take simple forms:
\<
E_{1}\eq 4\sin^2 (\pi sT/S),
\\\nn
E_{2,3}\eq
3+\sin^2(2\pi s' T/S) \pm \cos(2\pi s' T/S) \sqrt{9-8\sin^2(\pi sT/S)+\sin^2(2\pi s'T/S)}\,.
\label{eq:zero_charge}
\>
This is because the orbifold preserves a $\alg{su}(2)$ symmetry
(acting on the scalars $\phi_2$ and $\phi_3$)
which relates one of the three states with
the single-excitation state implying that
$E_1$ must coincide with \eqref{eq:singleex}.
Note that the only untwisted operator with two fields
and non-trivial anomalous dimension is an orbifold descendant of the Konishi operator
and inherits its anomalous dimension $E=6$.
Other choices of orbifold weights can also be analyzed, but they
lead to lengthy expressions for the anomalous dimensions which we will
not list here.

\section{Bethe Ansatz}

From the standpoint of the string theory dual it is
easy to argue that, in the strong 't~Hooft coupling limit, orbifold
field theories of $\superN=4$ SYM are integrable.
Assuming integrability, we shall investigate
the gauge theory Bethe ansatz for the orbifold model.
Let us first of all review integrability for
the parent $\superN=4$ supersymmetric model.
An eigenstate on a spin chain is determined by a
set of excitations.
There are eight types of excitations for the model, labeled
by $j=0,\ldots, 7$.
Excitations $j=1,\ldots,7$ correspond to spin waves which change
the flavors of spin sites. They are associated with a
spectral parameter $u_{j,k}$
where $k=1,\ldots K_j$ enumerates
the various excitations of type $j$.
The quasi-excitation $j=0$ corresponds to the insertion of
a new spin chain site. This type of excitation does not have
an associated spectral parameter, so it suffices to
give the number $K_0=L$, which represents the length of the spin chain.

\subsection{Equations for $\superN=4$ SYM}

The leading order (one-loop) Bethe equations for $\superN=4$ SYM read
\cite{Beisert:2003yb}
\[\label{eq:BetheN4}
\lrbrk{\frac{u_{j,k}-\ihalf V_j}{u_{j,k}+\ihalf V_j}}^L
\mathop{\prod_{j'=1}^J\prod_{k'=1}^{K_{j'}}}_{(j',k')\neq(j,k)}
\frac{u_{j,k}-u_{j',k'}+\ihalf M_{j,j'}}{u_{j,k}-u_{j',k'}-\ihalf M_{j,j'}}
=1,
\qquad
\prod_{j=1}^J
\prod_{k=1}^{K_j}
\frac{u_{j,k}+\ihalf V_j}{u_{j,k}-\ihalf V_j}
=1.
\]
where $J=7$ is the rank of $\alg{su}(2,2|4)$, $M_{j,j'}$ is the
symmetric Cartan matrix and $V_j$ are the labels
which specify the representation of spin sites.
The first equation is the Bethe equation for the excitation $k$ of type $j$.
It ensures that the spin chain state is periodic.
The second equation is the momentum constraint which ensures the
cyclicity of the state so that we can interpret it as
a single-trace local operator. It is useful to view the momentum
constraint as the Bethe equation for the quasi-excitations of type $0$:
Periodicity for spin sites is equivalent to vanishing momentum.
Although there are $L$ quasi-excitations of type $0$,
there is only one corresponding Bethe equation, because
all of these quasi-excitations are equivalent, they have no
spectral parameter which might distinguish them.
We can therefore summarize all the Bethe equations as
\[\label{eq:BetheGeneric}
\mathop{\prod_{j'=0}^J\prod_{k'=1}^{K_{j'}}}_{(j',k')\neq(j,k)}
S_{j,j'}(u_{j,k},u_{j',k'})
=1
\]
with the scattering phases $S_{j,j'}=1/S_{j',j}$ of the excitations
\[\label{eq:BethePhasesN4}
S_{j,j'}=
\frac{u_{j,k}-u_{j',k'}+\ihalf M_{j,j'}}{u_{j,k}-u_{j',k'}-\ihalf M_{j,j'}}\,,
\qquad
S_{j,0}=1/S_{0,j}=\frac{u_{j,k}-\ihalf V_j}{u_{j,k}+\ihalf V_j}\,,\qquad
S_{0,0}=1.
\]
The energy for a periodic eigenstate,
alias the planar anomalous dimension of a local operator,
is given by
\[\label{eq:BetheEnergy}
E=\sum_{j=0}^J\sum_{k=1}^{K_j}
e_j(u_{j,k}),\qquad
e_j(u_{j,k})=\frac{i}{u_{j,k}+\ihalf V_j}-\frac{i}{u_{j,k}-\ihalf V_j}\,,
\qquad
e_0=0.
\]
%

\subsection{Orbifolding the Bethe Ansatz}
\label{sec:BetheOrbi}

\begin{figure}\centering
\begin{minipage}{300pt}
\setlength{\unitlength}{1pt}%
\small\thicklines%
\begin{picture}(300,55)(-50,-10)
\put(-40,00){\circle{15}}%
\put(-40,15){\makebox(0,0)[b]{$-t_2$}}%
\put(  0,00){\circle{15}}%
\put(  0,00){\makebox(0,0){$-$}}%
\put(  0,15){\makebox(0,0)[b]{$0$}}%
\put(  7,00){\line(1,0){26}}%
\put( 40,00){\circle{15}}%
\put( 40,15){\makebox(0,0)[b]{$-t_1$}}%
\put( 47,00){\line(1,0){26}}%
\put( 80,00){\circle{15}}%
\put( 80,00){\makebox(0,0){$+$}}%
\put( 80,30){\makebox(0,0)[b]{$2t_1-t_2$}}%
\put( 87,00){\line(1,0){26}}%
\put(120,00){\circle{15}}%
\put(120,00){\makebox(0,0){$+$}}%
\put(120,15){\makebox(0,0)[b]{$2t_2-t_1-t_3$}}%
\put(127,00){\line(1,0){26}}%
\put(160,00){\circle{15}}%
\put(160,00){\makebox(0,0){$+$}}%
\put(160,30){\makebox(0,0)[b]{$2t_3-t_2$}}%
\put(167,00){\line(1,0){26}}%
\put(200,00){\circle{15}}%
\put(200,15){\makebox(0,0)[b]{$-t_3$}}%
\put(207,00){\line(1,0){26}}%
\put(240,00){\circle{15}}%
\put(240,00){\makebox(0,0){$-$}}%
\put(240,15){\makebox(0,0)[b]{$0$}}%
\put( 35,-5){\line(1, 1){10}}%
\put( 35, 5){\line(1,-1){10}}%
\put(195,-5){\line(1, 1){10}}%
\put(195, 5){\line(1,-1){10}}%
\end{picture}
\end{minipage}\vspace{0.5cm}

\begin{minipage}{300pt}
\setlength{\unitlength}{1pt}%
\small\thicklines%
\begin{picture}(300,55)(-50,-10)
\put(-40,00){\circle{15}}%
\put(-40,15){\makebox(0,0)[b]{$0$}}%
\put(  0,00){\circle{15}}%
\put(  0,00){\makebox(0,0){$+$}}%
\put(  0,15){\makebox(0,0)[b]{$0$}}%
\put(  7,00){\line(1,0){26}}%
\put( 40,00){\circle{15}}%
\put( 40,00){\makebox(0,0){$+$}}%
\put( 40,15){\makebox(0,0)[b]{$0$}}%
\put( 47,00){\line(1,0){26}}%
\put( 80,00){\circle{15}}%
\put( 80,00){\makebox(0,0){$+$}}%
\put( 80,15){\makebox(0,0)[b]{$0$}}%
\put( 87,00){\line(1,0){26}}%
\put(120,00){\circle{15}}%
\put(120,15){\makebox(0,0)[b]{$t_1$}}%
\put(127,00){\line(1,0){26}}%
\put(160,00){\circle{15}}%
\put(160,00){\makebox(0,0){$-$}}%
\put(160,30){\makebox(0,0)[b]{$t_2-2t_1$}}%
\put(167,00){\line(1,0){26}}%
\put(200,00){\circle{15}}%
\put(200,00){\makebox(0,0){$-$}}%
\put(200,15){\makebox(0,0)[b]{$t_1-2t_2+t_3$}}%
\put(207,00){\line(1,0){26}}%
\put(240,00){\circle{15}}%
\put(240,00){\makebox(0,0){$-$}}%
\put(240,30){\makebox(0,0)[b]{$t_2-2t_3$}}%
\put(115,-5){\line(1, 1){10}}%
\put(115, 5){\line(1,-1){10}}%
\end{picture}
\end{minipage}\vspace{0.5cm}

\begin{minipage}{300pt}
\setlength{\unitlength}{1pt}%
\small\thicklines%
\begin{picture}(300,55)(-50,-10)
\put(-40,00){\circle{15}}%
\put(-40,15){\makebox(0,0)[b]{$-t_2$}}%
\put(  0,00){\circle{15}}%
\put(  0,15){\makebox(0,0)[b]{$t_1$}}%
\put(  7,00){\line(1,0){26}}%
\put( 40,00){\circle{15}}%
\put( 40,00){\makebox(0,0){$-$}}%
\put( 40,15){\makebox(0,0)[b]{$0$}}%
\put( 47,00){\line(1,0){26}}%
\put( 80,00){\circle{15}}%
\put( 80,30){\makebox(0,0)[b]{$t_1-t_2$}}%
\put( 87,00){\line(1,0){26}}%
\put(120,00){\circle{15}}%
\put(120,00){\makebox(0,0){$+$}}%
\put(120,15){\makebox(0,0)[b]{$2t_2-t_1-t_3$}}%
\put(127,00){\line(1,0){26}}%
\put(160,00){\circle{15}}%
\put(160,30){\makebox(0,0)[b]{$t_3-t_2$}}%
\put(167,00){\line(1,0){26}}%
\put(200,00){\circle{15}}%
\put(200,00){\makebox(0,0){$-$}}%
\put(200,15){\makebox(0,0)[b]{$0$}}%
\put(207,00){\line(1,0){26}}%
\put(240,00){\circle{15}}%
\put(240,15){\makebox(0,0)[b]{$t_3$}}%
\put( -5,-5){\line(1, 1){10}}%
\put( -5, 5){\line(1,-1){10}}%
\put( 75,-5){\line(1, 1){10}}%
\put( 75, 5){\line(1,-1){10}}%
\put(155,-5){\line(1, 1){10}}%
\put(155, 5){\line(1,-1){10}}%
\put(235,-5){\line(1, 1){10}}%
\put(235, 5){\line(1,-1){10}}%
\end{picture}
\end{minipage}\vspace{0.5cm}

\caption{Orbifold weights for the simple roots
using the Dynkin diagrams ``Beauty{}'',
``Beast{}'' and ``Higher{}''
(from top to bottom).
The leftmost root represents a site of the spin chain,
a quasi-excitation of type $0$.
The indicated numbers are the orbifold weights $s_j$
for each type of Bethe root.}
\label{fig:Dynkin}
\end{figure}
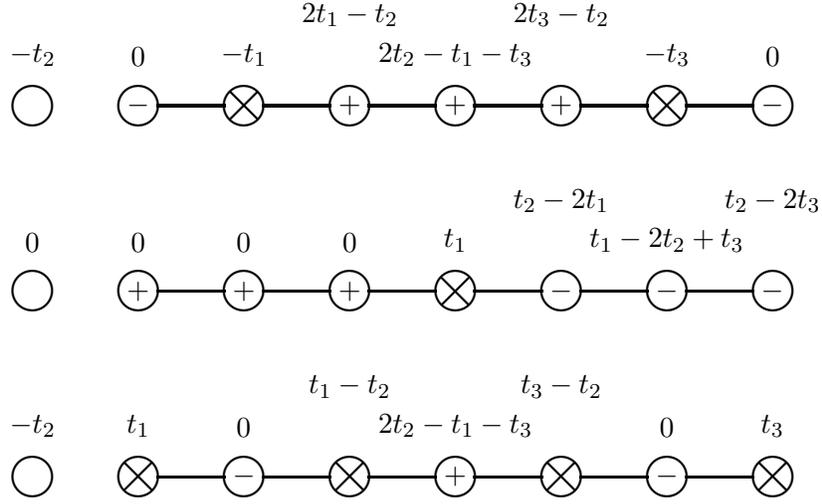

Keeping the above discussion in mind, the constraint \eqref{eq:OrbifoldShift}
suggests how to extend the Bethe ansatz to orbifolds:
This equation determines the phase shift
$2\pi s_X/S$ for exchanging a
flavored spin chain site $X$ with the $\Integers_S$ generator $\gamma$.
Therefore it is straightforward to represent $\gamma$ by a new type of
quasi-excitation,
i.e.~we introduce excitations of type $j=-1$ and set the lower bound
for $j'$ in \eqref{eq:BetheGeneric} to $j'=-1$.
Like the quasi-excitation with $j=0$ it does not
have an associated spectral parameter. The
only relevant parameter is $K_{-1}=T$ which
specifies the twisted sector.
The phase shift
\[\label{eq:BethePhasesOrbi}
S_{j,-1}=1/S_{-1,j}=\exp(2\pi i s_j/S)
\]
for exchanging $j'=-1$ with
any of the other $j=0,\ldots,7$ has to be
adjusted so that \eqref{eq:OrbifoldShift}
is respected.
This is simplified by the fact that
the phase shifts are determined by \eqref{eq:OrbifoldPhase}
through the $\alg{su}(4)$ charges.
Consequently, the numbers $s_{j}$
are defined by \eqref{eq:OrbifoldPhase}
through the charges $\vect{q}_j$ of simple roots of the algebra.
For superalgebras there are several
equivalent choices of simple roots,
and thus Dynkin diagrams and Cartan matrices $M_{j,j'}$.
There are three useful choices for $\alg{su}(2,2|4)$ which
we label by ``Beauty{}'', ``Beast{}'' and ``Higher{}'', see
\figref{fig:Dynkin}.
The numbers $s_j$, $j=(0\mathpunct{|}1,\ldots,7)$, for those three choices
are given by
\<\label{eq:BetheShifts}
\vect{s}\indup{Beauty}\eq(-t_2\mathpunct{|}0,-t_1,2t_1-t_2,2t_2-t_1-t_3,2t_3-t_2,-t_3,0),\nln
\vect{s}\indup{Beast}\eq(0\mathpunct{|}0,0,0,t_1,t_2-2t_1,t_1-2t_2+t_3,t_2-2t_3),\nln
\vect{s}\indup{Higher}\eq(-t_2\mathpunct{|}t_1,0,t_1-t_2,2t_2-t_1-t_3,t_3-t_2,0,t_3).
\>

Finally, the scattering phase of two $\gamma$'s is trivial because they commute.
They also do not contribute to the energy directly
\[\label{eq:BethePhasesOrbi2}
S_{-1,-1}=1,\qquad
e_{-1}=0.
\]

The one-loop Bethe equations for a $\Integers_S$ orbifold
of $\superN=4$ SYM are just as in
\eqref{eq:BetheGeneric}
\[\label{eq:BetheOrbiGen}
\mathop{\prod_{j'=-1}^J\prod_{k'=1}^{K_{j'}}}_{(j',k')\neq(j,k)}
S_{j,j'}(u_{j,k},u_{j',k'})
=1,\qquad
E=\sum_{j=-1}^J\sum_{k=1}^{K_j}
e_j(u_{j,k})
\]
but extended by the quasi-excitations of type $j,j'=-1$.
Written out explicitly, the Bethe equations read
\[\label{eq:BetheOrbi1}
e^{2\pi i T s_j/S}\lrbrk{\frac{u_{j,k}-\ihalf V_j}{u_{j,k}+\ihalf V_j}}^L
\mathop{\prod_{j'=1}^J\prod_{k'=1}^{K_{j'}}}_{(j',k')\neq(j,k)}
\frac{u_{j,k}-u_{j',k'}+\ihalf M_{j,j'}}{u_{j,k}-u_{j',k'}-\ihalf M_{j,j'}}
=1.
\]
Furthermore, there is the momentum constraint for $j=0$ 
\[\label{eq:BetheOrbi2}
e^{2\pi i T s_0/S}
\prod_{j'=1}^{J}
\prod_{k'=1}^{K_{j'}}
\frac{u_{j',k'}+\ihalf V_{j'}}{u_{j',k'}-\ihalf V_{j'}}
=1
\]
and the twist constraint
for $j=-1$
\[\label{eq:BetheOrbi3}
e^{-2\pi i L s_0/S}
\prod_{j'=1}^{J} e^{-2\pi i K_{j'} s_{j'}/S}=1.
\]
These equations agree with the
special cases studied in \cite{Wang:2003cu,Ideguchi:2004wm}.

The Bethe vacuum leading to these equations is $\Tr\gamma^T \fld{Z}^J$.
However, depending on the peculiarities of the orbifold group and its
chosen action, the twist constraint may project out the Bethe vacuum while
keeping some of the excited states. This is somewhat different from
the string theory orbifold construction in which, at least in a
Green-Schwarz formulation, the ground state is always part of the
spectrum. Since we assumed that the twisted sectors are decoupled, it
may be possible to construct one-loop Bethe equations whose form explicitly
depends on the twisted sector and length of the operators, such that
the Bethe vacuum is not projected out. However, the explicit
dependence on the length prevents a direct generalization to
higher loops. As we will see shortly in \secref{sec:Higher}, 
the setup outlined here is compatible with the 
higher-loop Bethe ansatz for $\superN=4$ SYM
and quantum strings \cite{Beisert:2005fw}.

Note that the modification of the Bethe equations 
is reminiscent of fractional magnetic flux
surrounded by the spin chain. 
In this picture, there are $T$ units of $1/S$ fractional 
flux associated to the $T$-twisted sector. 
The Bethe roots have electric charges $s_j$.
This means the particles undergo an Aharonov-Bohm phase shift
of $2\pi s_j T/S$ when moving once around the spin chain
as in \eqref{eq:BetheOrbi1,eq:BetheOrbi2}.
The twist constraint \eqref{eq:BetheOrbi2} 
represents the Dirac charge quantisation condition.
The twist matrix $\gamma$ as a quasi-excitation
can be viewed as the boundary of a coordinate patch
similar to Dirac strings or branch cuts. 
It does not carry a momentum and moving it
around corresponds to changing coordinate
patches by a gauge transformation.

\subsection{Energies}

Using these equations it is rather straightforward to recover the
one-loop anomalous dimensions computed before as well as some results
existing in the literature
\cite{Kim:2002fp,Ideguchi:2004wm,Wang:2003cu}. Quite obviously, the
simplest states to analyze are those with a single excitation of type
$j=4$ above the Bethe vacuum. Of course, such operators do not exist
in all 
theories, as they must satisfy the twist constraint \eqref{eq:BetheOrbi3}.
This is the equation \eqref{eq:condition} expressed in terms of Bethe
roots 
and for a single excitation it is equivalent to
\eqref{eq:singleexconstraint}. 
When this constraint is satisfied,
the (unique) rapidity is determined by the momentum constraint
\eqref{eq:BetheOrbi2} as
\[
u=-\frac{1}{2}\,\cot\frac{\pi s_0 T}{S}\,.
\]
Substituting this in \eqref{eq:BetheEnergy}
reproduces the anomalous dimension
\eqref{eq:singleex} obtained from
the explicit application of the dilatation operator.

Our second test concerns the operators whose anomalous dimensions
are listed in \eqref{eq:zero_charge}.
They correspond to length-two states of the spin chain
with two main excitations; for the ``Beauty{}''
Dynkin diagram, the number of excitations for each node is
\[
(L\mathpunct{|}K_1,K_2,K_3,K_4,K_5,K_6,K_7)=(2\mathpunct{|}0, 0,1,2,1,0,0)~.
\]
According to \eqref{eq:OrbifoldSU3} the choice
$\vect{s}_{\rep{3}}=(2s',s,s)$
is related $\vect{t}=(s-s',s,s+s')$
which again implies the weights
$\vect{s}\indup{Beauty}=(-s\mathpunct{|}0,-s+s',s-2s',0,s+2s',-s-s',0)$
for Bethe roots.
The equations \eqref{eq:BetheOrbi1} can be solved explicitly
as solutions to a bi-quadratic equation.
Evaluating \eqref{eq:BetheEnergy} on these solutions leads to the anomalous
dimensions \eqref{eq:zero_charge}.

Last, but not least, it is easy to recover the anomalous dimensions of
BMN-type operators with excitations of a single type, originally
discussed in \cite{Kim:2002fp,Ideguchi:2004wm,Wang:2003cu}. In this
case only the  orbifold weight corresponding to $j=4$
(in the ``Beauty{}'' form) is relevant.
Then, assuming that the number of excitations is
such that the operator twist constraint is satisfied, the logarithm of
the Bethe equations implies that
\[
u_{4,k}=\frac{L}{2\pi(n_k+Ts_4/S)}~~~~~\mbox{with}~~~~\sum_{k=1}^K n_k=0.
\]
With these rapidities, the equation \eqref{eq:BetheEnergy} yields
the known result
\[
E=\frac{4\pi^2}{L^2}\sum_{k=1}^K \lrbrk{n_k+\frac{Ts_4}{S}}^2.
\]
%

\subsection{Specific Orbifolds}

Let us investigate the orbifold parameters for several special cases.
The first is an orbifold preserving $\superN=1$ superconformal
symmetry.
This is easily achieved by demanding that
in the ``Beast{}'' representation
\eqref{eq:BetheShifts},
the fermionic excitation $j=4$ commutes with
the $\Integers_S$ generator $j'=-1$.
In other words,
we should not not only have unbroken
$\alg{su}(2,2)$ ($s_1=s_2=s_3=0$ is guaranteed)
but also unbroken
$\alg{su}(2,2|1)$, i.e.~$s_4=0$.
This means we set $t_1=0$ and obtain
for a generic $\superN=1$ orbifold
\<\label{eq:ModelsN1}
\vect{s}_{\rep{4}}\eq (0,-t_2,t_2-t_3,t_3),\qquad
\vect{s}_{\rep{3}}=(t_3,t_3-t_2,t_2),\nln
\vect{s}\indup{Beauty}\eq(-t_2\mathpunct{|}0,0,-t_2,2t_2-t_3,2t_3-t_2,-t_3,0),\nln
\vect{s}\indup{Beast}\eq(0\mathpunct{|}0,0,0,0,t_2,t_3-2t_2,t_2-2t_3),\nln
\vect{s}\indup{Higher}\eq(-t_2\mathpunct{|}0,0,-t_2,2t_2-t_3,t_3-t_2,0,t_3).
\>
To go to a $\superN=2$ orbifold we need to have $s_4=s_5=0$ in the Beast
representation, i.e.~$t_1=t_2=0$. Without loss of generality we can then set
$t_3=1$, i.e.~$\vect{t}=(0,0,1)$, and obtain
\<\label{eq:ModelsN2}
\vect{s}_{\rep{4}}\eq(\phantom{+}0,\phantom{+}0,-1,+1),\qquad
\vect{s}_{\rep{3}}=(+1,+1,\phantom{+}0),\nln
\vect{s}\indup{Beauty}\eq(\phantom{+}0\mathpunct{|}\phantom{+}0,\phantom{+}0,\phantom{+}0,-1,+2,-1,\phantom{+}0),\nln
\vect{s}\indup{Beast}\eq(\phantom{+}0\mathpunct{|}\phantom{+}0,\phantom{+}0,\phantom{+}0,\phantom{+}0,\phantom{+}0,+1,-2),\nln
\vect{s}\indup{Higher}\eq(\phantom{+}0\mathpunct{|}\phantom{+}0,\phantom{+}0,\phantom{+}0,-1,+1,\phantom{+}0,+1).
\>
For $S=2$ the symmetry is enhanced by $\alg{su}(2)$.

We could also demand a preserved $\alg{su}(3)$ symmetry.
This is achieved for $s_5=s_6=0$ or $6t_1=3t_2=2t_3$.
Without loss of generality we set $\vect{t}=(1,2,3)$
and obtain
\<\label{eq:ModelsSU3}
\vect{s}_{\rep{4}}\eq (-1,-1,-1,+3),\qquad
\vect{s}_{\rep{3}}=(+2,+2,+2),\nln
\vect{s}\indup{Beauty}\eq(-2\mathpunct{|}\phantom{+}0,-1,\phantom{+}0,\phantom{+}0,+4,-3,\phantom{+}0),\nln
\vect{s}\indup{Beast}\eq(\phantom{+}0\mathpunct{|}\phantom{+}0,\phantom{+}0,\phantom{+}0,+1,\phantom{+}0,\phantom{+}0,-4),\nln
\vect{s}\indup{Higher}\eq(-2\mathpunct{|}+1,\phantom{+}0,-1,\phantom{+}0,+1,\phantom{+}0,+3).
\>
The cases $S=2,4$ are special, they are the only orbifolds with
preserved $\alg{su}(4)$ symmetry.
For $S=3$ we have $\superN=1$ superconformal symmetry instead.

Another interesting option is to preserve $\alg{su}(2)\times\alg{su}(2)$ symmetry.
This is achieved for $s_5=s_7=0$ or $t_2=2t_3=2t_1$.
Without loss of generality we set $\vect{t}=(1,2,1)$
and the weights in various representations read
\<\label{eq:ModelsSU22}
\vect{s}_{\rep{4}}\eq (-1,-1,+1,+1),\qquad
\vect{s}_{\rep{3}}=(\phantom{+}0,\phantom{+}0,+2),\nln
\vect{s}\indup{Beauty}\eq(-2\mathpunct{|}\phantom{+}0,-1,\phantom{+}0,+2,\phantom{+}0,-1,\phantom{+}0),\nln
\vect{s}\indup{Beast}\eq(\phantom{+}0\mathpunct{|}\phantom{+}0,\phantom{+}0,\phantom{+}0,+1,\phantom{+}0,-2,\phantom{+}0),\nln
\vect{s}\indup{Higher}\eq(-2\mathpunct{|}+1,\phantom{+}0,-1,+2,-1,\phantom{+}0,+1).
\>
%

\subsection{Higher Loops, Quantum Strings}
\label{sec:Higher}

We would now like to see whether the orbifold is compatible with the Bethe
ansatz for higher-loop $\superN=4$ SYM \cite{Beisert:2005fw}.
These equations seem to be equally well applicable to 
quantum strings on $AdS_5\times S^5$ when 
introducing an additional overall phase 
for the strong coupling regime \cite{Arutyunov:2004vx}.
In these ans\"atze one uses the rapidities $x_{j,k}$ instead of $u_{j,k}$
and the scattering phases
$S_{j,j'}(x_{j,k},x_{j',k'})$
between the various kinds of Bethe
roots are modified to accommodate the higher-loop interactions, 
cf.~\cite{Beisert:2005fw} for full expressions.
The generic form of the Bethe equations, however,
remains unchanged as compared to \eqref{eq:BetheGeneric}.
In these Bethe equations we cannot use any Dynkin diagram,
but we should restrict to the one we denoted by ``Higher{}''
in \figref{fig:Dynkin}.
Another very important ingredient for the consistency of the equations
is the ``dynamic transformation{}'' \cite{Beisert:2005fw}
which ensures compatibility with the dynamic spin chains
that arise at higher loops \cite{Beisert:2003ys}.
This transformation is the result of a symmetry of the scattering
phases
\[\label{eq:HigherDynamic}
S_{j,3}(x,x_3)=S_{j,1}(x,x_1)\,S_{j,0}(x),\qquad
S_{j,5}(x,x_5)=S_{j,7}(x,x_7)\,S_{j,0}(x)
\]
when $x_1x_3=x_5x_7=\gym^2 N/16\pi^2$.
The scattering phases for $\superN=4$ SYM with $j=0,\ldots 7$ obey this
symmetry. In order for the orbifold to be compatible with the
transformation, the equation \eqref{eq:HigherDynamic}
should hold for $j=-1$ as well:
\[
e^{2\pi i s_3/S}=e^{2\pi i s_1/S} e^{2\pi i s_0/S},\qquad
e^{2\pi i s_5/S}=e^{2\pi i s_7/S} e^{2\pi i s_0/S}.
\]
This means that $s_1+s_0=s_3$ and $s_5=s_7+s_0$
in the ``Higher{}'' choice of Dynkin diagram, which is indeed
satisfied by \eqref{eq:BetheShifts}.
Therefore the higher-loop Bethe equations are consistent
with the orbifolding procedure and
\eqref{eq:BetheOrbiGen} will most likely
produce equally accurate results for orbifolds
as the original equations
\eqref{eq:BetheGeneric}
for plain $\superN=4$ SYM.%
\footnote{
As we have mentioned before, due to the quasi-excitation representing
the orbifold group element, the spin chain Hamiltonian for
finite-length operators potentially differs from the one of
$\superN=4$ SYM one loop order before the effects of wrapping
interactions become relevant. It is therefore possible that the prediction
of the Bethe ansatz for the anomalous dimensions of length $L$ operators
depart from their actual anomalous dimensions at $(L-1)$ loops.}

\section{Generalizations}

In general, the number of deformations of a field theory is quite
large. Deformations preserving certain properties of the parent
theory -- such as integrability for example -- are less
numerous. If a theory exhibits at least one global $\grp{U}(1)$ symmetry it
is always possible to construct its abelian orbifold descendants
which, following the arguments described above, exhibit an integrable
dilatation operator. If the parent theory has a string theory dual,
a similar conclusion can be reached in the strong coupling regime. 

Our discussion naturally generalizes to arbitrary abelian
orbifolds. Since all group generators commute, they can be
simultaneously diagonalized and each of them spans a copy of
$\Integers_S$ for some $S$. 
Thus, a general abelian orbifold is isomorphic to an orbifold by
a product of $\Integers_{S_j}$ factors, $\otimes_{j=1}^P \Integers_{S_j}$,
and independent orbifold weights $\vect{t}_j$ for each factor.
The operators are organized in (twisted) sectors with respect 
to each factor of the orbifold group, which are labelled by $T_j$.
Then starting from \eqref{eq:BetheOrbi1,eq:BetheOrbi2}, the
relevant Bethe equations are then obtained via the following
replacements: $S\rightarrow S_{j'}$, $T\rightarrow T_{j'}$, $s_{j'}\rightarrow s_{j',j}$.
The index $j'=1,\ldots,P$ runs over all the factors in the orbifold group. 
The equation \eqref{eq:BetheOrbi3} is replaced by a system of equations
enforcing the fact that the eigenoperators are built out of fields
invariant under each factor of the orbifold group.

Another direction for generalizations are so-called $\beta$-deformations.
If a theory exhibits several commuting $\grp{U}(1)$ symmetries it is 
possible to construct its $\beta$-deformation 
\cite{Roiban:2003dw,Berenstein:2004ys,Lunin:2005jy,Frolov:2005ty,Frolov:2005dj}. 
It was argued in \cite{Frolov:2005ty,Beisert:2005if} 
that this deformation preserves integrability in
the weak coupling regime. If the parent theory has a string theory
dual, then the dual of the deformed theory can also be constructed
\cite{Lunin:2005jy} and it can be argued that integrability is
preserved also in the strong coupling regime
\cite{Frolov:2005ty}. From this description it
is clear that it is possible to combine these two
deformations. Moreover, as both deformations preserve the existing
$\grp{U}(1)$ symmetries, the order in which the
deformations are performed is inconsequential. If the parent theory is
$\superN=4$ SYM, the Bethe equations
\eqref{eq:BetheOrbi1,eq:BetheOrbi2,eq:BetheOrbi3} are 
modified by adjoining
to the left hand side the $\beta$-dependent phase deformations
discussed in the section 4 of \cite{Beisert:2005if}.
In short, an antisymmetric matrix $A_{j,j'}$ is
introduced which gives an additional constant phase for commuting one
Bethe root of type $j$ past a Bethe root of type $j'$.

We now assemble the above two generalizations with the 
higher-loop equations of \secref{sec:Higher}:
The generic higher-loop equations
for deformed, multiply orbifolded $\superN=4$ SYM read
($j=1,\ldots,7$, $k=1,\ldots,K_j$)
\[
\prod_{j'=1}^P
e^{2\pi i T_{j'} s_{j',j}/S_{j'}}
\,\,\,
e^{iA_{j,0}L}
\prod_{j'=1}^7
e^{iA_{j,j'}K_{j'}}
\,\,\,
S_{j,0}(x_{j,k})^L
\mathop{\prod_{j'=1}^7\prod_{k'=1}^{K_{j'}}}_{(j',k')\neq(j,k)}
S_{j,j'}(x_{j,k},x_{j',k'})
=1.
\]
Furthermore, there is the momentum constraint
\[
\prod_{j'=1}^P
e^{2\pi i T_{j'} s_{j',0}/S_{j'}}
\prod_{j'=1}^7
e^{iA_{0,j'}K_{j'}}
\prod_{j'=1}^{7}
\prod_{k'=1}^{K_{j'}}
S_{0,j'}(x_{j',k'})
=1,
\]
and the twist constraint ($j=1,\ldots,P$)
\[
e^{-2\pi i s_{j,0} L/S_j}
\prod_{j'=1}^{7} e^{-2\pi i  s_{j,j'}K_{j'}/S_{j}}=1.
\]
Consistency of the dynamic transformation 
\eqref{eq:HigherDynamic} leads to the following restrictions
\[\begin{array}[b]{rclcrcl}
A_{j,0}+A_{j,1}\eq A_{j,3},&&
s_{j,0}+s_{j,1}\eq s_{j,3},\\[3pt]
A_{j,0}+A_{j,7}\eq A_{j,5},&&
s_{j,0}+s_{j,7}\eq s_{j,5}.
\end{array}
\]
Note that this constraint is even more general than
the \secref{sec:BetheOrbi}: It also includes the possibility
to orbifold and deform parts of the $AdS_5$ space. 

When assigning the $P$ twist matrices $\gamma_j$
as quasi-excitations $j=-1,\ldots,-P$
we can also write the Bethe equations in short as
\[
\mathop{\prod_{j'=-P}^7\prod_{k'=1}^{K_{j'}}}_{(j',k')\neq(j,k)}
S^A_{j,j'}(x_{j,k},x_{j',k'})
=1.
\]
with suitably twisted scattering phases $S^A_{j,j'}$ for
between all types of excitations $j,j'=1,\ldots,7$,
spin chain sites $j,j'=0$ and twist matrices
$j,j'=-1,\ldots,-P$.

It is worth
emphasizing that, while both the $\beta$-deformation and the orbifold
projection lead to similar modifications to the Bethe equations, they
are in fact quite different. In physical terms, the difference
is similar to that between assigning (non-commuting) magnetic charges to
all spin states
vs.~having a system with fractional magnetic flux (the orbifold field theory),
cf.~\secref{sec:BetheOrbi}. 
If the number of excitations is small compared to the length of the chain
(the BMN limit)
the two physical pictures become similar, as the large number of
fields constituting the BMN vacuum may be interpreted as a 
twist matrix provided its charge is identified with the orbifold twist.
This is reflected by the similarity of the anomalous dimensions
in this limit, which are given by the original BMN spectrum
with mode numbers shifted by a fixed fractional amount.

The quasi-excitation representing the orbifold group element has a
physical interpretation closely related to string theory. On the
orbifold covering space, twisted strings are interpreted as open
strings in which the world sheet fields the two ends are related by
the action of the orbifold group element. 
Similarly here, it is possible to interpret the spin chain for 
the orbifold theory as an open spin chain with boundary
conditions specified by an orbifold group element.%
\footnote{Some integrable open spin chains with more general boundary
conditions are  
also connected to the one-loop dilatation operators for theories with
fields in the fundamental representation
\cite{Chen:2004mu,DeWolfe:2004zt,Erler:2005nr,Berenstein:2005vf,McLoughlin:2005gj}.}
The quasi-excitation representing the orbifold group element is an
effective way of implementing this boundary condition while
keeping the spin chain closed, similarly with the twisted boundary
conditions for closed strings on orbifolds.

\section{Discussions and Outlook}

In this paper we have discussed in detail the integrability of
orbifolds of four-dimensional field theories with integrable
dilatation operators, with emphasis on $\superN=4$ SYM. We showed
how to modify the Bethe equations to incorporate an abelian orbifold
construction and performed a number of simple checks on the resulting
system. It should not be complicated to compare the Bethe ansatz
predictions with the classical sigma model on
$AdS_5\times S^5/\orbi$.
For that one would have to implement the orbifold
projection at the level of the spectral curve of the sigma model on
$AdS_5\times S^5$.

Our discussion focused on abelian orbifolds. From a field theory
standpoint it is possible to construct non-abelian orbifold projections
as well. It would be interesting to understand the consequences of
such a projection on the integrable structure of the parent theory.
For such theories the twisted sectors are associated to the  conjugacy
classes of the orbifold group and, consequently, the analog of 
\eqref{eq:BetheOrbi3} becomes more
involved. Moreover, it is likely to be necessary to include more than
one quasi-excitation which also exhibit non-trivial
scattering phases among themselves.
It might also be worth considering more general 
theories with $\superN<4$ supersymmetry 
\cite{Ferretti:2004ba,DiVecchia:2004jw,Belitsky:2004cz,Beisert:2004fv}
such as quiver gauge theories \cite{Douglas:1996sw}
and model with fundamental matter 
\cite{Chen:2004mu,DeWolfe:2004zt,Erler:2005nr,Berenstein:2005vf,McLoughlin:2005gj}.

It would also be interesting to understand the details of the gauge
and string  
theory description of orbifolds of $AdS_5$. As we have already mentioned, 
the from the standpoint of the Bethe ansatz it merely requires
allowing the non-trivial orbifold weights $\vect{s}$ corresponding to
the roots of the conformal group, the details of the field theory
leading to such an ansatz (even at one loop) are not completely clear.

A very interesting problem is understanding the effects on the
anomalous dimensions of the double-trace deformations induced at loop
level \cite{Tseytlin:1999ii,Dymarsky:2005uh,Dymarsky:2005nc}
in the absence of supersymmetry. As we have already mentioned, 
for length $L$ operators these deformations cannot have an
effect of order one before the $L-1$ loop order. From this perspective
their effects are not unlike those of the wrapping interactions and
understanding how  to incorporate the former in the Bethe ansatz may
shed light on the latter.

\paragraph{Acknowledgments.}

We would like to thank Didina Serban and Mathias Staudacher for
discussions.
R.~R.~is grateful for hospitality at KITP and Princeton University
where parts of this work was carried out.
R.~R.~acknowledges partial support of NSF grant PHY99-07949 while at KITP.
The work of N.~B.~is supported in part by the U.S.~National Science
Foundation Grant No.~PHY02-43680.
Any opinions, findings and conclusions or recommendations expressed in this
material are those of the authors and do not necessarily reflect the
views of the National Science Foundation.

\bibliographystyle{nb}
\bibliography{ZMorbi}

\end{document}